# Stochastic Rounding for Image Interpolation and Scan Conversion

Olivier Rukundo, Samuel Emil Schmidt

Department of Health Science and Technology
Faculty of Medicine, Aalborg University
Aalborg, Denmark

*Abstract*—The stochastic rounding (SR) function is proposed to evaluate and demonstrate the effects of stochastically rounding row and column subscripts in image interpolation and scan conversion. The proposed SR function is based on a pseudorandom number, enabling the pseudorandom rounding up or down any non-integer row and column subscripts. Also, the SR function exceptionally enables rounding up any possible cases of subscript inputs that are inferior to a pseudorandom number. The algorithm of interest is the nearest-neighbor interpolation (NNI) which is traditionally based on the deterministic rounding (DR) function. Experimental simulation results are provided to demonstrate the performance of NNI-SR and NNI-DR algorithms before and after applying smoothing and sharpening filters of interest. Additional results are also provided to demonstrate the performance of NNI-SR and NNI-DR interpolated scan conversion algorithms in cardiac ultrasound videos.

*Keywords—Cardiac ultrasound; deterministic rounding; image quality; interpolation; pseudorandom number; scan conversion; stochastic rounding; video quality*

## I. INTRODUCTION

Image interpolation is an important type of estimation that pervades many engineering applications, where estimates of image pixel values at points other than the input or source grid are required and/or affect the desired results and/or the way to obtain them [1]. In digital image upscaling, the nearest neighbor interpolation (NNI) remains the fastest algorithm. NNI is used for estimating image pixel values at points of interest, and it is traditionally based on the deterministic rounding (DR) function. This type of function deterministically rounds off the subscripts of the grid coordinates of the output or destination image to enable the mapping of pixels from the source image into the destination image. However, in some NNI interpolated images cases, the image quality is often bad because of the presence of heavy jagged artefacts, especially at image objects' edges. Here, the DR function-based mapping remains main the inherent flaws that contribute to the presence of such heavy jagged artefacts. In this work, the stochastic rounding (SR) function is alternatively proposed to demonstrate and evaluate the effects or benefits of stochastically rounding row and column subscripts in NNI-based image interpolation and scan conversion. Note that, the scan conversion algorithm is used for translating input data (captured in different coordinates) into Cartesian coordinates (still more suitable for display) [9]. More information on the scan conversion system block diagram and the scan conversion using bilinear interpolation examples are provided in [9], [10]. It is important to note that, when the fractional part of non-integer subscripts equals half a unity, this, key challenge in rounding functions, still raises the question of how or when to reasonably use the gain or loss of half a unity – and the SR function is the answer to this question. The rest of the paper is organized as follows: Section two introduces the literature review of interest. Section three presents the SR function. Section four presents numerical examples showing the comparison of results based on SR and DR functions. Section five presents experimental results. Relevant discussions are provided in Section six. The conclusion is given in Section seven.

## II. LITERATURE REVIEW

There exist many image interpolation algorithms, in various categories, that were developed focusing on improving the accuracy or efficiency of the algorithm, depending on targeted applications including but not limited to rescaling, reslicing, rendering, zooming, coordinate transformations in two-dimensional data or scan conversion, tomographic reconstruction and image registration [1]. One of the most recent applications of interest is artificial intelligence (AI)-based image super-resolution [2], [3] - whose generalizable steps are shown in Fig. 1. Here, the authors' core idea was to enhance the quality of the bilinear interpolation images, by applying a set of pre-learned filters on the image patches, chosen by an efficient hashing mechanism [2]. In another example, presented in [3], the authors used the bicubic image interpolation algorithm to upscale the input or source image to meet the same size as the reference image before starting to recover from it, a resolution enhanced image comparable to the ground truth high-resolution image. Both bilinear and bicubic interpolation algorithms belong to the extra pixel category [4]. This category encompasses all image interpolation algorithms that create non-original pixels to achieve interpolation results [4]. The NNI algorithm belongs to the non-extra pixel category [4]. This category of image interpolation algorithms does not create non-original pixels to achieve interpolation results [5]. As mentioned earlier, the NNI algorithm remains the fastest among image interpolation algorithms [6], [7]. However, the NNI algorithm is based on the deterministic rounding for source image pixels selection. In [4], the author demonstrated that the best deterministic rounding (DR) function for nearest neighbor image interpolation purposes was the ceil function. In [8], authors proposed the stochastic rounding (SR) idea and, according to authors in [8], the SR idea was attracting renewed interest in artificial intelligence/deep learning because it could improve the accuracy of the underlying computations.

Here, for x∈R with x∉F (where F⊆R denotes the floating-point number system), the authors considered two stochastic rounding modes shown in Eq. 1 and Eq. 2.

Mode-1:

$$fl(x) = \begin{cases} \lceil x \rceil \text{ with probability } 0.5 \\ \lfloor x \rfloor \text{ with probability } 0.5 \end{cases} \quad (1)$$

Mode-2:

$$fl(x) = \begin{cases} \lceil x \rceil \text{ with probability } p = (x - \lfloor x \rfloor)/(\lceil x \rceil - \lfloor x \rfloor) \\ \lfloor x \rfloor \text{ with probability } 1 - p \end{cases} \quad (2)$$

In the first mode (or Eq. 1) authors round x∈R with x∉F up or down with equal probability to the respective nearest floating-point number. In the second mode (or Eq. 2), authors round with a probability that is 1 minus the relative distance of x to each of the nearest floating-point numbers. For x∈R, $\lfloor x \rfloor$ = max {y∈F: y⩽x}, $\lceil x \rceil$ = min {y∈F: y⩾x}, so that $\lfloor x \rfloor$ ⩽x ⩽$\lceil x \rceil$ with equality throughout if x∈F. For x∉F, $\lfloor x \rfloor$ and $\lceil x \rceil$ are adjacent floating-point numbers. More details are provided in [8].

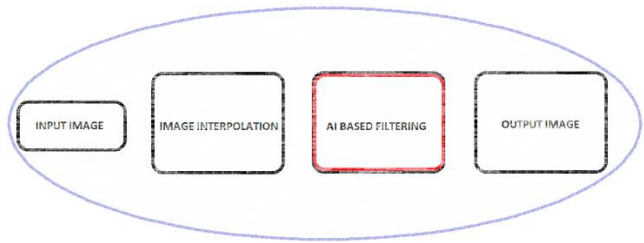

Fig. 1. An Example of Generalizable Steps for AI-based Image Super-Resolution Application Pervaded by Image Interpolation.

### III. STOCHASTIC ROUNDING FUNCTION

Here, the stochastic rounding function is developed based on the equation that incorporates a rand function found in MATLAB. The MATLAB rand function is based on a new pseudorandom number generator named Mersenne Twister (MT). According to [11], the MT pseudorandom number generator seems to be the best among all generators ever implemented, with the period $2^{19937} − 1$ and 623-dimensional equidistributional property. Also, the success of this C-Code MT19937 has been achieved thanks to two new ideas added to the previous version, Generalized Feedback Shift Register (GFSR), namely, (1) the incomplete array, and (2) the inversive-decimation method [11]. In the experimental simulations, presented in this work, the pseudorandom number (r) was rounded to one digit. Also, the pseudorandom number was tuned to randomly vary between 0 and 0.5. In this way, it was possible to automate probabilities of stochastically rounding up or down thus achieving non-zero positive integers to be used as row and column subscripts of pixel coordinates. Note that, due to the intended application - of rounding non-integer row and column subscripts - Eq.3 incorporates conditions that allow it to only output non-zero positive integers.

$$sr(x) = \begin{cases} \lceil x - r \rceil, if\ x > r \wedge (x\ mod\ 1) > 0 \\ \lceil x \rceil, otherwise \end{cases} \quad (3)$$

In this way, the first condition ensures that any input index or subscript is greater than any pseudorandom number varying between 0 and 0.5. The second condition ensures that any input subscript is of non-integer type before proceeding to randomly rounding up or down. Note that, for 0⩽r⩽0.5, it is an exception if r > x. Therefore, in such an exceptional case, the SR function rounds up (e.g., see Table I: See the first line in the 4X group).

Fig. 2 shows an example of destination pixel coordinates subscripts before round-off operations. In Fig. 2(a) and (b) results were obtained by doubling a 3-by-3 matrix and plotting the coordinates of the matrix elements. Note that, when the rounding operation is random - this results in stochastic pixel selection in NNI.

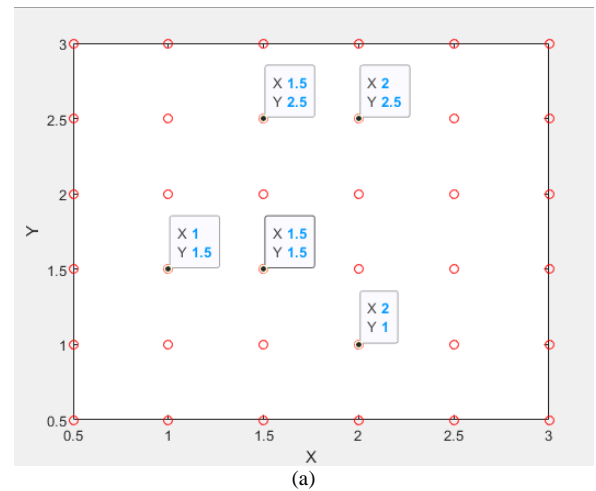

(a)

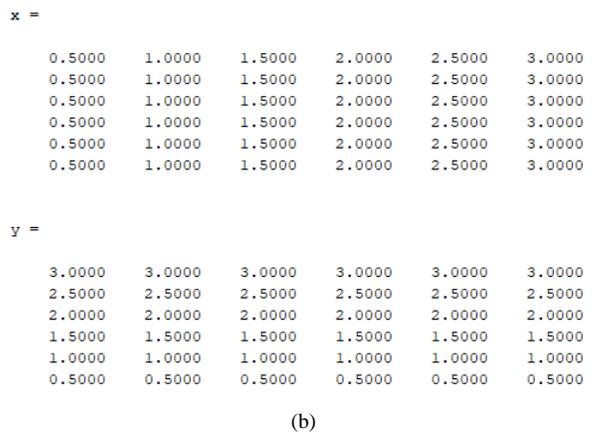

(b)

Fig. 2. (a) Shows Pixel Coordinate Subscripts before Round-off Operations. (b) Shows Row (y) and Column (x) Subscripts before Round-off Operations.

### IV. NUMERICAL EXAMPLES

In Table I, the authors compare the output of SR and DR functions - after upscaling the original 3-by-3 matrix, two times, three times, and four times. As can be seen, Table I shows a column of subscripts, a column of random numbers, a column of DR results, and a column of SR results as well as columns of the elapsed time in both cases. Again, in Table I, when the scaling ratio is equal to two, the SR results differ from the DR results, twice. The same happens when the scaling ratio is equal to three. When the scaling ratio is equal to four, the SR results differ from the DR results, three times, except that in this case, there is an exception, mentioned earlier - about when x<r.

TABLE I. X REPRESENTS SUBSCRIPTS AND r REPRESENTS A RANDOM NUMBER (ROUNDED TO ONE DIGIT). THE DR(X) IS EQUIVALENT TO THE CEIL(X) FUNCTION

| ratio | x | r | DR($x$) | SR($x$) | DR (*sec*) | SR (*sec*) |
|---|---|---|---|---|---|---|
| 2X | 0.5000 | 0.4 | 1 | 1 | 0.18 × 1.0e-05 | 0.68 × 1.0e-05 |
|  | 1.0000 | 0.5 | 1 | 1 |  |  |
|  | **1.5000** | **0.5** | **2** | **1** |  |  |
|  | 2.0000 | 0.1 | 2 | 2 |  |  |
|  | **2.5000** | **0.5** | **3** | **2** |  |  |
|  | 3.0000 | 0.3 | 3 | 3 |  |  |
| 3X | 0.3333 | 0.1 | 1 | 1 | 0.15 × 1.0e-05 | 0.60 × 1.0e-05 |
|  | 0.6667 | 0.4 | 1 | 1 |  |  |
|  | 1.0000 | 0.3 | 1 | 1 |  |  |
|  | **1.3333** | **0.5** | **2** | **1** |  |  |
|  | 1.6667 | 0.3 | 2 | 2 |  |  |
|  | 2.0000 | 0.2 | 2 | 2 |  |  |
|  | **2.3333** | **0.4** | **3** | **2** |  |  |
|  | 2.6667 | 0.3 | 3 | 3 |  |  |
|  | 3.0000 | 0.1 | 3 | 3 |  |  |
| 4X | **0.2500** | **0.5** | 1 | 1 | 0.04 × 1.0e-05 | 0.17 × 1.0e-05 |
|  | 0.5000 | 0.1 | 1 | 1 |  |  |
|  | 0.7500 | 0.4 | 1 | 1 |  |  |
|  | 1.0000 | 0 | 1 | 1 |  |  |
|  | 1.2500 | 0.4 | **2** | **1** |  |  |
|  | **1.5000** | **0.4** | 2 | 2 |  |  |
|  | 1.7500 | 0.4 | 2 | 2 |  |  |
|  | 2.0000 | 0.4 | 2 | 2 |  |  |
|  | 2.2500 | 0.5 | **3** | **2** |  |  |
|  | **2.5000** | **0.1** | 3 | 3 |  |  |
|  | 2.7500 | 0.3 | 3 | 3 |  |  |
|  | 3.0000 | 0.4 | 3 | 3 |  |  |
|  | 3.2500 | 0.4 | **4** | **3** |  |  |
|  | **3.5000** | **0.4** | 4 | 4 |  |  |
|  | 3.7500 | 0.5 | 4 | 4 |  |  |
|  | 4.0000 | 0.1 | 4 | 4 |  |  |

Note that, SR may also produce similar results to DR results, but that is not guaranteed because the SR's output relies on the pseudorandom value. Also, it is important to note that, the results presented in Table I are specific to a particular case of r value and input numbers. Still, in Table I, it can be seen, in the first case involving 2X (1.5 and 2.5), the SR behaved like the floor function, instead of the traditional otherwise. In the second case involving 4X (1.5, 2.5, and 3.5), the SR behaved like the ceil function – in this way, the pseudo-randomness of the SR function has answered the question or removed the challenge of how to reasonably round a non-integer subscript in when the fractional part is equal to half a unity (i.e., 0.5). Note that, the most interesting point of DR and SR functions is that, when the fraction part of a non-integer equals 0.5, the DR function always rounds a non-integer in a predetermined or deterministic way, which is not the case with the SR function. Now, comparing the elapsed time or line reading time, it can be seen, in Table I, that the time taken by both the SR and DR functions (to round a given series of non-integers) is too small to make any significant difference.

## V. EXPERIMENTS

### A. Datasets, Smoothing / Sharpening Method, IQA Metrics

*1) Dataset*: Here, the used image dataset originated from the USC-SIPI Database of 210 Textures, Aerials, Miscellaneous, and Sequences images [12]. Here the author uses input images of 128 ×128 size and reference images of 512 x 512 size, all converted to 8bits using R2020a MATLAB. All experimental images are also available at the author's GitHub via GitHub.com/orukundo [13].

*2) Smoothing / sharpening method*: The 2-D Gaussian smoothing kernel and sharpened using the unsharp masking methods - available in the MATLAB 2020a image processing toolbox - are used to extend experiments via evaluating smoothed and sharpened interpolation results.

*3) IQA metrics*: In the beginning, only full-reference (FR) IQA metrics are used. Those included the mean-squared error (MSE), structural similarity index (SSIM), and peak signal to noise ratio (PSNR). These FR-IQA metrics are selected to quantify or measure the closeness or similarity of modified or distorted images (i.e., in this case, interpolated images) against their corresponding pristine images (i.e., reference images), [14]. Note that for SSIM and PSNR, normally when the scores are higher (closer to 1 and 100) that means the better visual quality. For MSE when the scores are lower (closer to 0), that normally means better visual quality. Here, it is important to note that MATLAB's *tic* and *toc* command function is also used to check the elapsed time while reading code lines of the SR and DR functions (as shown in Table I). In the end, - given that there exist no reference images or videos for cardiac ultrasound images or videos - the video frames quality assessment is done using no-reference (NR) IQA metrics. The selected NR-IQA metric of interest is the Perception-based Image Quality Evaluator (PIQE) [15], [16]. Specifically, PIQE is used to calculate one frame's no-reference perceptual image quality after every 78-milliseconds for 10 000 milliseconds (i.e., entire video duration). This 78 milliseconds timestamp is estimated based on the number of frames of each video and the entire video duration as well as the suitability for graphical representation. To understand the PIQE scores, the quality scale, and score range are as follows: Excellent [0 ↔ 20]. Good [21 ↔ 35]. Fair [36 ↔ 50]. Poor [51 ↔ 80]. Bad [81 ↔ 100], [16]. For scan conversion operations, T5D data files are used after being acquired from Duke University's Experimental Ultrasound System, T5, [10], [17]. More information on Duke University's Experimental Ultrasound System, T5 can be found via [18],[19],[20]. Note that, with T5D files, scan conversion operations are doable using a dedicated graphical user interface, developed in MATLAB, for post-processing of those ultrasound image sequences (from Duke University's Experimental Ultrasound System, T5).

*B. Automatic / Objective Evaluation Results (Natural Images)*

In Table II and Table III, priority is given to NNI-DR and NNI-SR results over the bicubic and bilinear results – keeping in mind that traditional bicubic and bilinear algorithms generally perform much better than the traditional NNI algorithm.

TABLE II. SSIM, PSNR AND MSE METRICS SCORE ESTIMATES BEFORE APPLYING SMOOTHING AND SHARPENING FILTERS)

|  | SSIM | | PSNR | | MSE | |
|---|---|---|---|---|---|---|
|  | NNI-DR | NNI-SR | NNI-DR | NNI-SR | NNI-DR | NNI-SR |
| IMAGE1 | 0.4809 | 0.5127 | 21.387 | 22.058 | 472.42 | 404.82 |
| IMAGE2 | 0.5306 | 0.5688 | 21.443 | 22.391 | 466.41 | 374.34 |
| IMAGE3 | 0.8713 | 0.8825 | 30.174 | 31.395 | 62.469 | 47.152 |
| IMAGE4 | 0.8230 | 0.8388 | 23.667 | 24.903 | 279.46 | 210.24 |
| IMAGE5 | 0.5555 | 0.5859 | 20.186 | 21.162 | 622.88 | 497.50 |
| IMAGE6 | 0.5067 | 0.5480 | 20.606 | 21.586 | 565.45 | 451.30 |

TABLE III. SSIM, PSNR AND MSE METRICS SCORE ESTIMATES AFTER APPLYING SMOOTHING AND SHARPENING FILTERS

|  | SSIM | | PSNR | | MSE | |
|---|---|---|---|---|---|---|
|  | NNI-DR | NNI-SR | NNI-DR | NNI-SR | NNI-DR | NNI-SR |
| IMAGE1 | 0.5112 | 0.5481 | 21.381 | 22.334 | 473.03 | 379.89 |
| IMAGE2 | 0.5627 | 0.6081 | 21.338 | 22.658 | 477.77 | 352.58 |
| IMAGE3 | 0.8857 | 0.9050 | 29.910 | 31.727 | 66.38 | 43.68 |
| IMAGE4 | 0.8358 | 0.8637 | 23.221 | 25.070 | 309.67 | 202.32 |
| IMAGE5 | 0.5876 | 0.6354 | 20.009 | 21.528 | 648.86 | 457.30 |
| IMAGE6 | 0.5333 | 0.5895 | 20.516 | 21.977 | 577.32 | 412.45 |

*C. Subjective / Human Evaluation Results (Natural Images)*

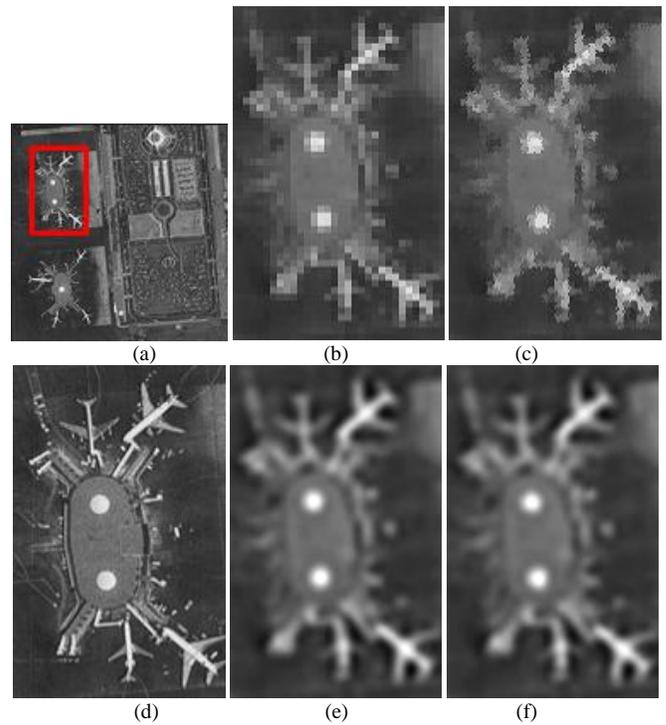

Fig. 3. (a) Input Image1. (b) NNI-DR Interpolated Image1. (c) NNI-SR Interpolated Image1. (d) RF Image1. (e) (b)-Filtered. (f) (c)-Filtered.

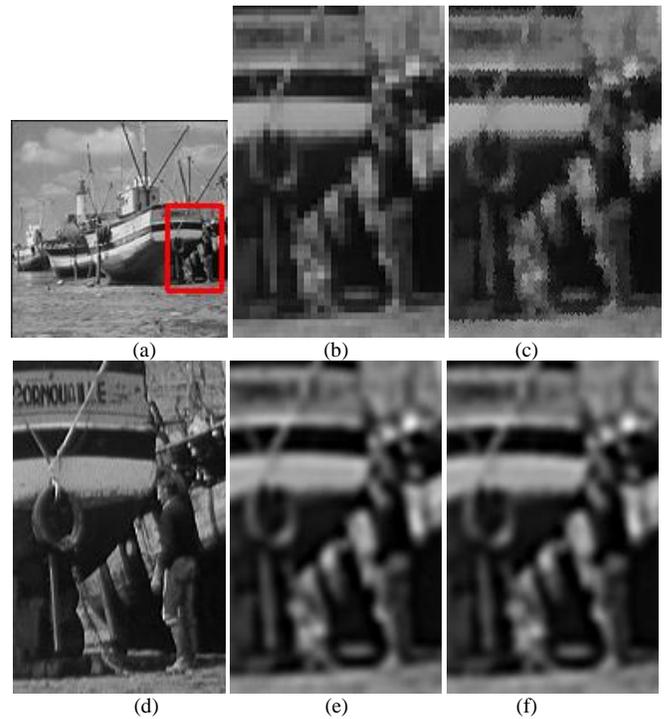

Fig. 4. (a) Input Image2. (b) NNI-DR Interpolated Image2. (c) NNI-SR Interpolated Image2. (d) RF Image2. (e) (b)-Filtered. (f) (c)-Filtered.

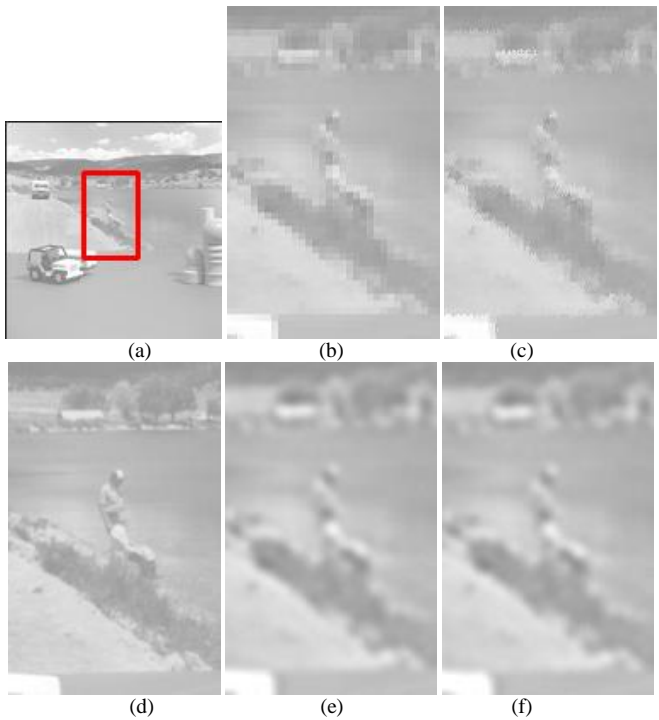

Fig. 5. (a) Input Image3. (b) NNI-DR Interpolated Image3. (c) NNI-SR Interpolated Image3. (d) RF Image3. (e) (b)-Filtered. (f) (c)-Filtered.

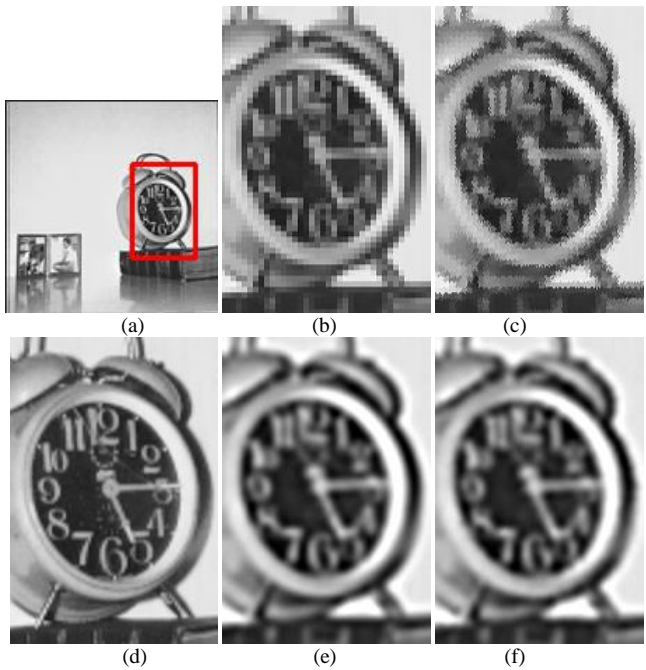

Fig. 6. (a) Input Image4. (b) NNI-DR Interpolated Image4. (c) NNI-SR Interpolated Image4. (d) RF Image4. (e) (b)-Filtered. (f) (c)-Filtered.

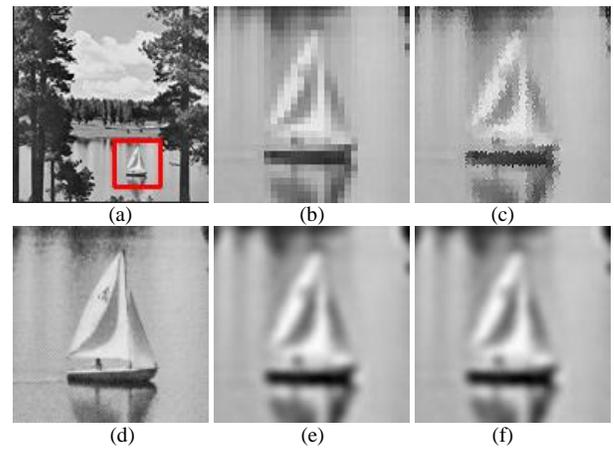

Fig. 7. (a) Input Image5. (b) NNI-DR Interpolated Image5. (c) NNI-SR Interpolated Image5. (d) RF Image5. (e) (b)-Filtered. (f) (c)-Filtered.

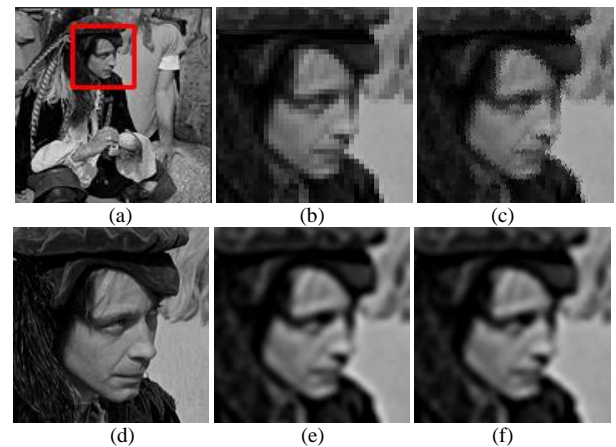

Fig. 8. (a) Input Image6. (b) NNI-DR Interpolated Image6. (c) NNI-SR Interpolated Image6. (d) RF Image6. (e) (b)-Filtered. (f) (c)-Filtered.

*D. Subjective / Human Evaluation Results (Sectored Images)*

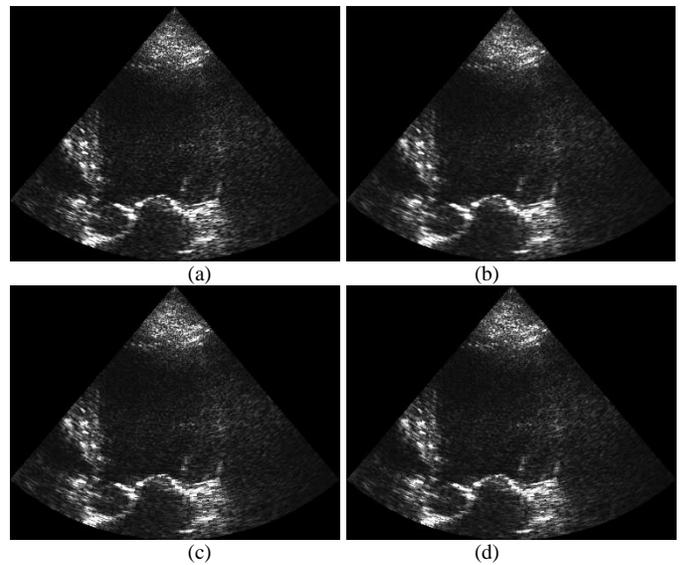

Fig. 9. (a) Bicubic, (b) Bilinear, (c) NNI-DR, (d) NNI-SR (Also see [36]: Interpolated Scan Conversion - 60).

## VI. Discussion

Table II shows columns of score estimates achieved by NNI-DR and NNI-SR algorithms (using the SSIM, PSNR, and MSE FR-IQA metrics). In all five image cases presented, the NNI-SR algorithm achieved score estimates slightly higher than the NNI-DR score estimates relevant to SSIM and PSNR. Also, in all five image cases, the NNI-SR algorithm achieved score estimates slightly lower than the NNI-DR score estimates relevant to MSE. The same situation is repeated in Table III, where, unlike in Table III, the results presented are achieved after applying the smoothing and sharpening filters. The Table II comparison concludes that the NNI-SR performed better than the NNI-DR, in this specific situation. However, from the author's observation, it is also possible that the slight betterment of the NNI-SR may have been caused by the fact that the content of NNI-SR images was better aligned with the content of the reference images than the content of the NNI-DR images.

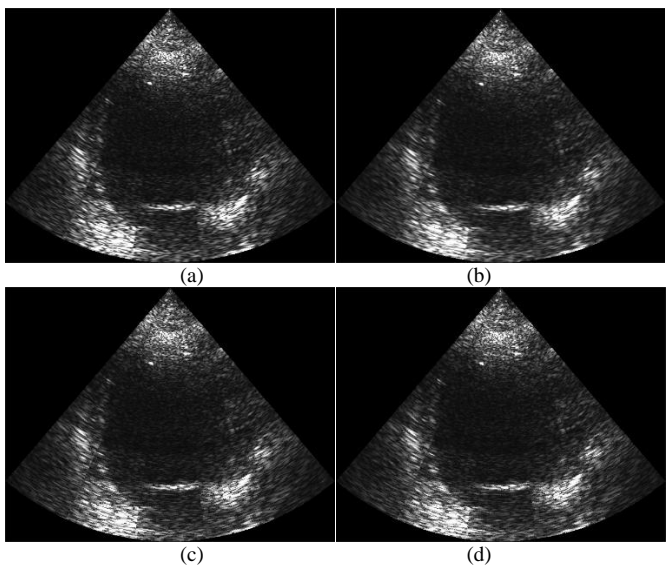

Fig. 10. (a) Bicubic, (b) Bilinear, (c) NNI-DR, (d) NNI-SR (Also see [36]: Interpolated Scan Conversion - 1000).

In Fig. 3, (b) and (c) images have both produced jagged edges on the outline of the planes as well as the rest of the building. Also, comparing the results achieved by NNI-SR and NNI-DR against the (d) RF image, it is clear that only the NNI-SR algorithm reconstructed the white dots in a way almost similar to the way such dots look in (d) image. Also, it is clear, the NNI-DR changed the three white dots or circles to squares. A similar situation did not repeat after smoothing and sharpening the results of NNI-DR and NNI-SR algorithms, shown in (e) and (f), respectively. Note that, here, getting a closer and clear view of the planes (and/or their locations) was not possible, even if the results could show plans and airport terminal (even without having previously seen the RF image in (d)). In Fig. 4, (b) and (c) images have shown jagged edges on the outline of the standing man as well as the rest of the ship edges. As can be seen, the NNI-DR algorithm produced so heavy jagged artefacts that the silhouette of the man disappeared completely, as shown in (b). Now, with the image produced by the NNI-SR algorithm, in (c), the man's silhouette is only less hardly imaginable than in the NNI-DR case, shown in (b). After smoothing and sharpening, the results became too blurred that is impossible to imagine the man's silhouette, as shown in (e) and (f). If one sees the (e) and (f) images without having previously seen the RF image in (d), it is not possible to imagine the presence of a man's silhouette or ship edges. Here, it is also important to note that by cropping a small part of the image (a), the aim was to get a closer and clear view of the man. In Fig. 5(b) and (c) images have both produced heavy jagged edges on the outline of the two men as well as the rest of the beach. This demonstrates that getting a closer and clear view of the two men was not possible using both NNI-DR and NNI-SR. But, after smoothing and sharpening the (b) and (c) images, the results shown in (e) and (f) allow one to imagine the presence of two men, one standing and one sitting, without even having previously seen the RF image in (d). A similar situation is repeated in Fig. 6, Fig. 7, and Fig. 8.

Fig. 9 and Fig. 10 show two cardiac ultrasound imaging sectored images obtained using different interpolated scan conversion algorithms. Here, the bicubic interpolated scan conversion sectored image-(a) looks better than the rest of bilinear, NNI-DR, and NNI-SR interpolated scan conversion images in (b), (c), and (d), respectively. It is important to note that originally the frame rate was 60 and 1074 frames per second for images in Fig. 9 and Fig. 10, respectively. The value of frame rates could be seen on the Graphical User Interface developed for cardiac ultrasound video visualization. Note that relevant videos are available at https://github.com/orukundo/Interpolated-Scan-Conversion-of-B-Mode-Cardiac-Ultrasound-Image-Sequences (see the link entitled: Interpolated Scan Conversion-1000 and/or Interpolated Scan Conversion-60). Also, note that to easily perceive the video quality difference – relevant to the mentioned interpolated scan conversion algorithms - the monitor resolution must be high enough.

Further assessments were done via plotting and comparing graphs of pixel intensity distributions in NNI-DR and NNI-SR interpolated images against the pixel intensity distribution in reference images before and after filtering operations. Fig. 11(a) shows the number of pixels counts versus the corresponding number of bins in NNI-DR interpolated and unfiltered IMAGE1, NNI-SR interpolated and unfiltered IMAGE1, and RF IMAGE1. Fig. 11(b) shows the number of pixels counts versus the corresponding number of bins in NNI-DR interpolated and filtered IMAGE1, NNI-SR interpolated and filtered IMAGE1 and RF IMAGE1. As can be seen, in each case, none of the NNI-DR or NNI-SR results (represented by the blue and green line) matched perfectly with the RF results (represented by a red line). In other words, the number of pixels belonging to each bin of the FR IMAGE1 remained different from the number of pixels belonging to each bin of the NNI-DR and NNI-SR IMAGE1. This difference is also visible between NNI-DR and NNI-SR results, otherwise, it would not be possible to see the blue and green lines. Although not exactly at the same extent, a similar situation is generally repeated in Fig. 12, Fig. 13, Fig. 14, Fig. 15, and Fig. 16 based on IMAGE2, IMAGE3, IMAGE4, IMAGE5, and IMAGE6. Note that the number of empty bins in the source image remained equal to the number of empty bins in the images interpolated by NNI-DR and NNI-SR (i.e., a condition in the non-extra pixel category).

In Fig. 17(a) and (b) cases, the video frames were repeatedly evaluated after 78 milliseconds (instead of evaluating every frame), in each of the 10-seconds videos. This option for evaluation of video frames was opted to better understand the performance of each interpolated scan conversion algorithm, or else understand why an image or frame quality was bad or good at a specific time, in ultrasound systems. Here, it is important to note that, in the past, many attempts were done to develop methods to assess the quality of ultrasound imaging systems automatically or objectively [21], [22], [23]. In the recent past, the author introduced an index for image interpolation quality assessment as a preliminary step to a suitable method for image quality assessment in ultrasound imaging – only focusing on undesirable artefacts, known as aliasing [24].

the bilinear interpolation has always been associated with being the most blurriness productive among all non-adaptive interpolation algorithms [25], [26], [27], [28] even if it has proved to be useful in other image processing techniques [29], [30], [31]. The fact that bilinear performed poorer than others in both (a) and (b) cases, confirms its inherent flaw of being the most interpolation blurriness productive.

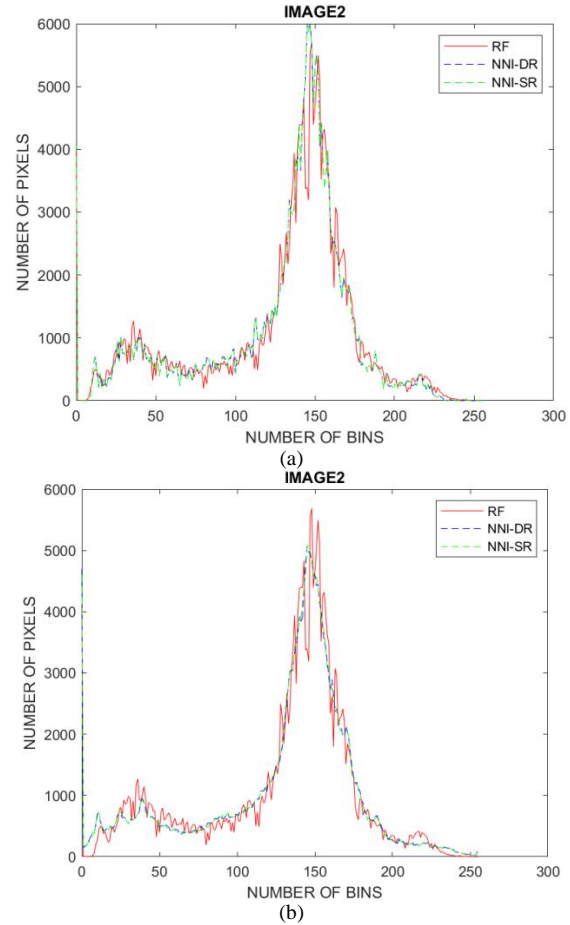

Fig. 12. (a) and (b) show the Number of Pixels Counts versus the Number of Bins in RF, NNI-DR and NNI-SR-based IMAGE2 before and after Filtering.

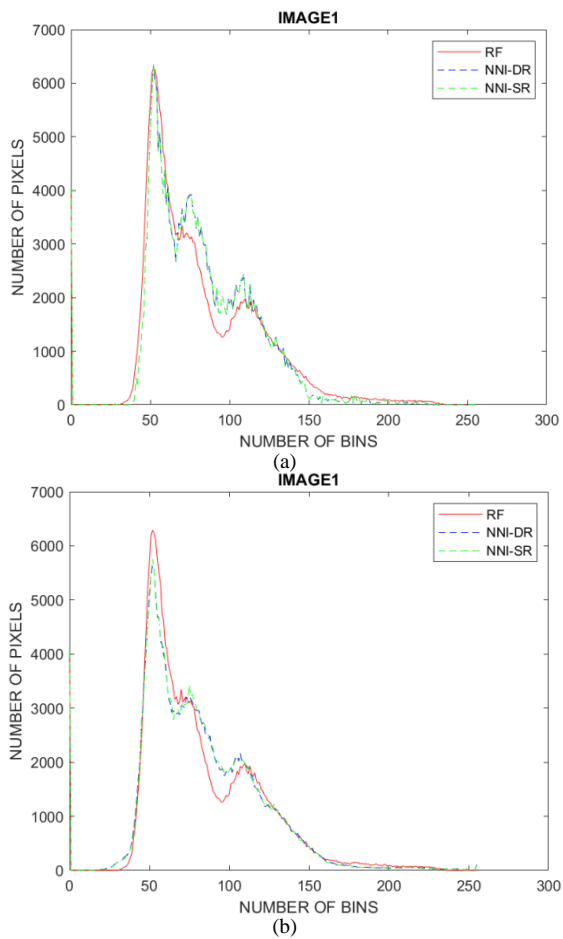

Fig. 11. (a) and (b) show the Number of Pixels Counts versus the Number of Bins in RF, NNI-DR and NNI-SR-based IMAGE1 before and after Filtering.

However, until now, the PIQE remains the only NR-IQA metric very sensitive to two interesting cases of undesirable artefacts in ultrasound imaging, namely: grain-like or speckle-like noise and blurriness. As can be seen, the bilinear interpolated scan conversion algorithm achieved the lowest mean score (i.e., blurriest video frames) in both (a) and (b) cases. Also, note that the current literature demonstrates that

Now, considering (b), where the original video frame rate was 60, NNI-SR achieved higher PIQE scores than NNI-DR and in general, both NNI-DR and NNI-SR interpolated scan conversion algorithms achieved the best PIQE scores compared to the other two interpolation algorithms of the extra-pixel category. Considering (a), where the original video frame rate was 1074, the bicubic interpolated scan conversion algorithm demonstrated strength that would normally be expected, as it normally produces better image quality than most non-adaptive interpolation algorithms [32], [33], [34], [35]. The mean scores were provided, in the legend, to quickly assess the performance of each interpolated scan conversion algorithm. Note that, these mean scores are specific to these cases. Also, note that these mean score values may change. It is important to note that, the 78 milliseconds timestamp was adopted referring to the frame rate and video duration to enable the more informative and clearer plotting of graphs – otherwise with only 10 seconds videos, the graphs would have looked like straight lines or similar. Note that the results in (a) and (b) also prove that the PIQE can be considered as the most suitable NR-IQA metric for ultrasound image quality assessment because, in (a) and (b) example, PIQE-based results match perfectly with the well-known performances of interpolation methods, mentioned.

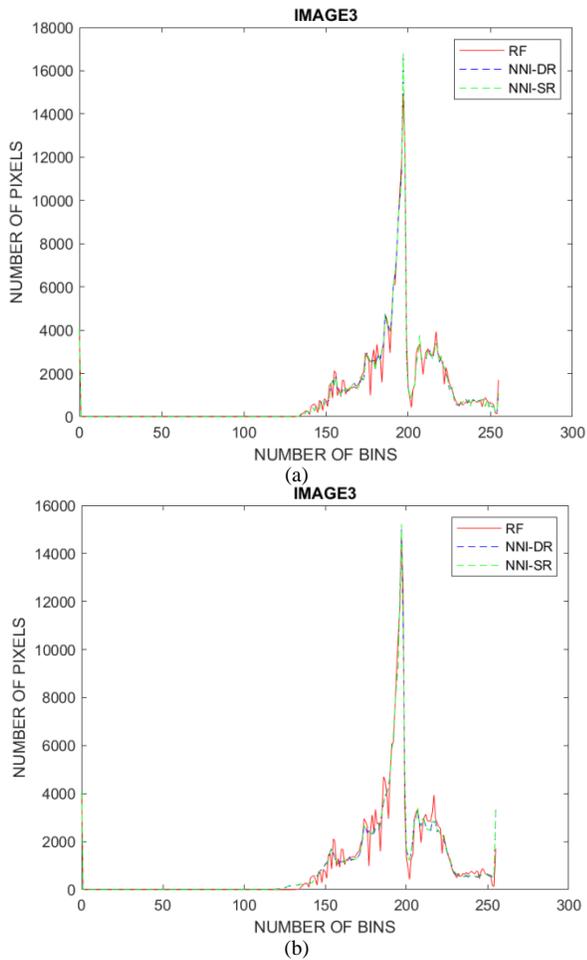

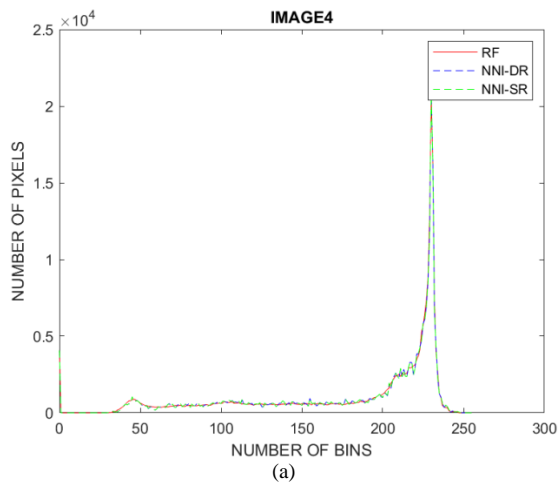

Fig. 13. (a) and (b) show the Number of Pixels Counts versus the Number of Bins in RF, NNI-DR and NNI-SR-based IMAGE3 before and after Filtering.

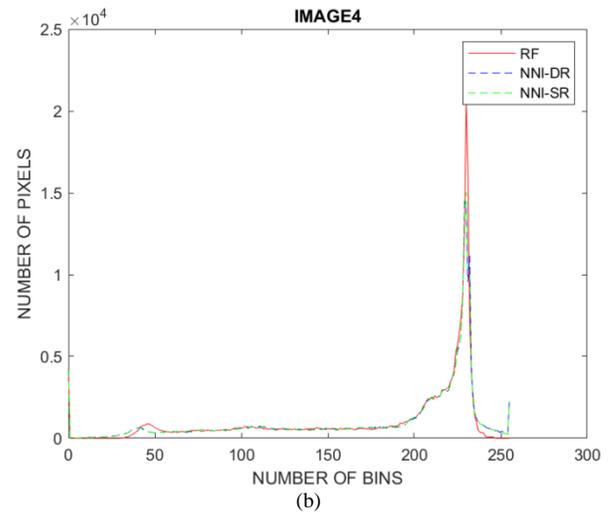

Fig. 14. (a) and (b) show the Number of Pixels Counts versus the Number of Bins in RF, NNI-DR and NNI-SR-based IMAGE4 before and after Filtering.

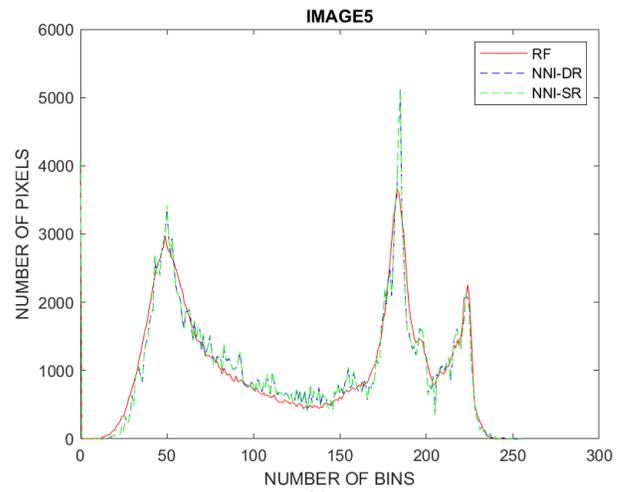

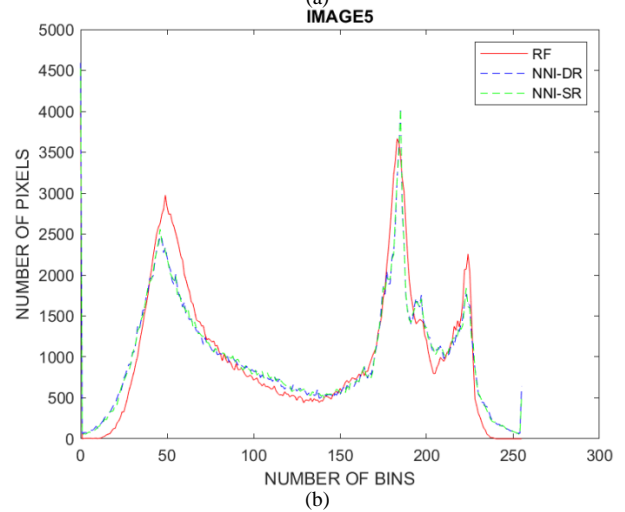

Fig. 15. (a) and (b) show the Number of Pixels Counts versus the Number of Bins in RF, NNI-DR and NNI-SR-based IMAGE5 before and after Filtering.

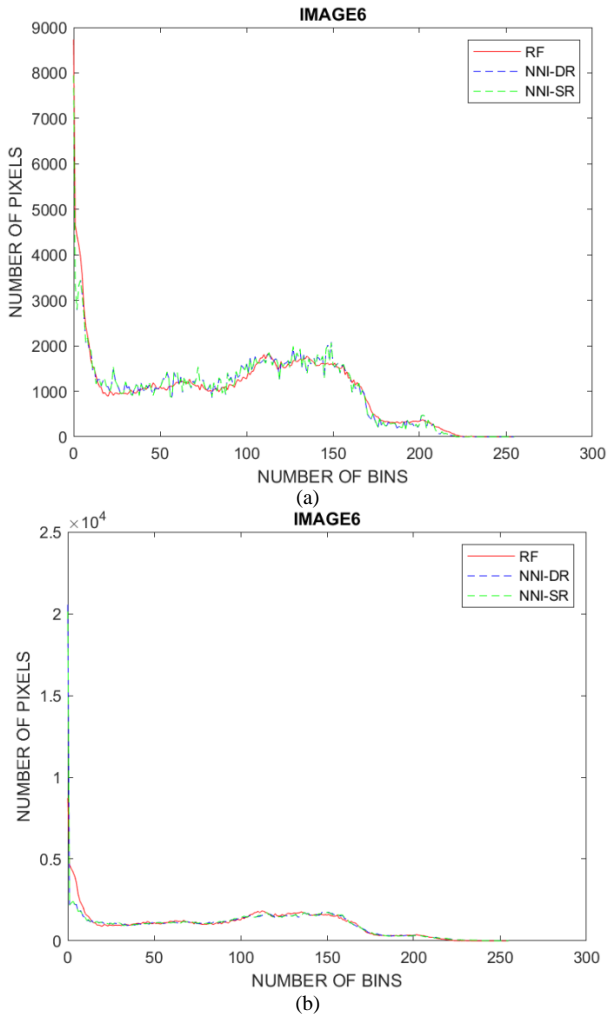

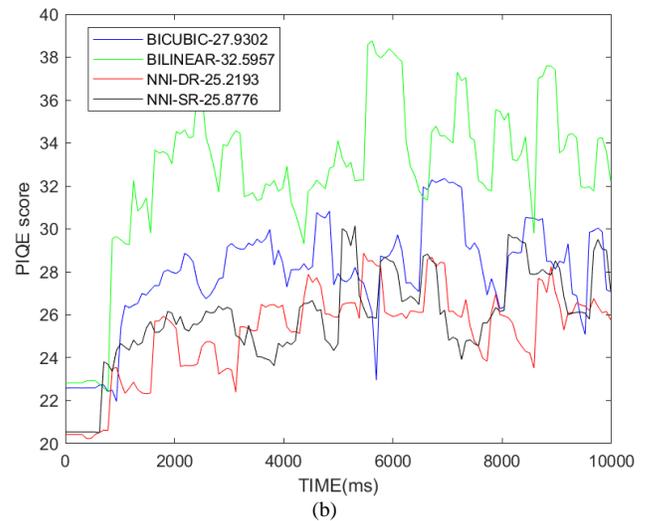

Fig. 16. (a) and (b) show the Number of Pixels Counts versus the Number of Bins in RF, NNI-DR and NNI-SR-based IMAGE6 before and after Filtering.

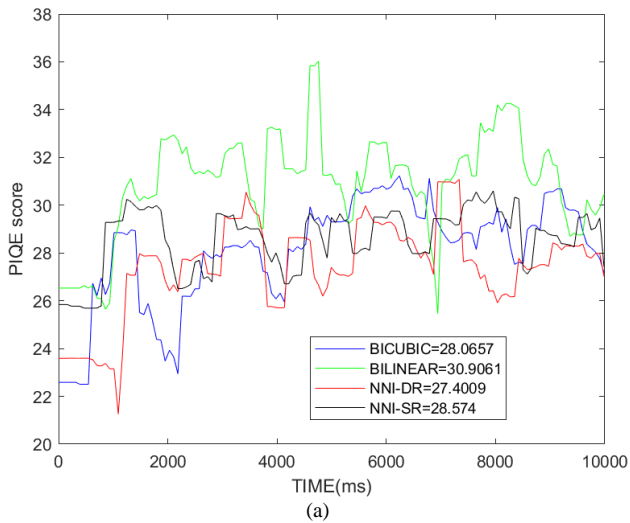

Fig. 17. (a): Video1 - Original Frame Rate Equals 1074. (b) Video2 - Original Frame Rate Equals 60.

## VII. CONCLUSION

Evaluation and demonstration of effects of stochastically rounding row and column subscripts in NNI-based image interpolation and interpolated scan conversion were presented and discussed. Here, evaluation of effects of SR function was achieved using both human evaluation and automatic IQA metrics of interest while experimental demonstrations were conducted using natural and ultrasound images. With natural images, the automatic evaluation showed that the NNI-SR algorithm could achieve slightly better score estimates than the NNI-DR score estimates in terms of SSIM, PSNR and MSE - before and after applying the smoothing and sharpening filters. However, the human evaluation showed that both rounding functions could result in heavy jagged artefacts at the edges of image objects, with the exception after the smoothing and sharpening of interpolated images. With cardiac ultrasound images, the human evaluation suggested that the bicubic interpolated scan conversion algorithm could achieve the best results among the rest of the algorithms (in these specific cases involving sectored images). With 60-fps and 1074 fps videos, the NNI-SR and NNI-DR almost tied in terms of PIQE scores thus raising the question of which algorithm could perform better than the other if longer videos were used. It is important to remind that relevant research challenges and extensive findings were presented and explained especially in the introduction and discussion parts. Future works could focus on applying the SR function in other engineering areas – for example, in a data augmentation to create more deep learning training samples, etc.

## VIII. COMPETING INTEREST

The authors declare that they have no competing interests.

## ACKNOWLEDGMENT

The authors would like to thank the reviewers and editors for their constructive comments. The author would like to thank Duke University, Biomedical Engineering Department for support.

REFERENCES

[1] Kumar, A., Agarwal, N., et al, An efficient 2-D jacobian iteration modeling for image interpolation, 2012 19th IEEE International Conference on Electronics, Circuits, and Systems, pp. 977-980, (2012).

[2] Romano, Y., et al., RAISR: Rapid and Accurate Image Super-Resolution, In: ArXiv:1606.01299v3, (2016).

[3] Dong, C., Loy, C.C., He, K., Tang, X., Learning a Deep Convolutional Network for Image Super-Resolution, In: Fleet D., Pajdla T., Schiele B., Tuytelaars T. (eds) Computer Vision, ECCV, (2014).

[4] Rukundo, O., Non-extra Pixel Interpolation, International Journal of Image and Graphics, 20(4), 2050031, (2020).

[5] Rukundo, O., Evaluation of Rounding Functions in Nearest Neighbour Interpolation, International Journal of Computational Methods, 10 pages, (2021).

[6] Rukundo, O., Cao, H.Q., Nearest Neighbour Value Interpolation, International Journal of Advanced Computer Science and Applications, 3(4), (2012).

[7] Rukundo, O., Maharaj, B.T., Optimization of Image Interpolation based on Nearest Neighbour Algorithm, International Conference on Computer Vision Theory and Applications, Lisbon, 641–647, (2014).

[8] Connolly, M., Higham, N., Mary, T., Stochastic Rounding and its Probabilistic Backward Error Analysis, SIAM Journal on Scientific Computing, Society for Industrial and Applied Mathematics, 43 (1), A566–A585, (2021).

[9] Li, X.H., Ultrasound Scan Conversion on TI's C64x+ DSPs, TI Application Report, SPRAB32, (2009).

[10] Rukundo, O., Schmidt, S.E., Von Ramm, O.T., Software Implementation of Optimized Bicubic Interpolated Scan Conversion in Echocardiography, In: ArXiv: 2005.11269, (2020).

[11] Matsumoto, M. Nishimura, T., Mersenne Twister: A 623-dimensionally equidistributed uniform pseudorandom number generator, ACM Transactions on Modeling and Computer Simulation, 8(1), 3–30, (1998).

[12] USC-SIPI Image Database: <http://sipi.usc.edu/database/database.php>, Accessed: 2021-10-12.

[13] Modified-USC-SIPI-Image-Database: <https://github.com/orukundo/Modified-USC-SIPI-Image-Database>, Accessed: 2021-10-12.

[14] Rukundo, O., Schmidt, S., Extrapolation for Image Interpolation, Proc. SPIE 10817, Optoelectronic Imaging and Multimedia Technology V, 108171F, (2018).

[15] Venkatanath, N., Praneeth, D., et al., Blind Image Quality Evaluation Using Perception Based Features, In Proceedings of the 21st National Conference on Communications (NCC). Piscataway, NJ: IEEE, (2015).

[16] Sheikh, H. R., Wang, Z., et al., LIVE Image Quality Assessment Database Release 2, <https://live.ece.utexas.edu/research/quality>, Accessed: 2021-10-12.

[17] Moore, C., Castellucci, J., et al., Live High-Frame-Rate Echocardiography, IEEE Transactions on Ultrasonics, Ferroelectrics, and Frequency Control, 62(10), 1779-1787, (2015).

[18] Andersen, M.V, Moore, C., et al., Quantitative Parameters of High-Frame-Rate Strain in Patients with Echocardiographically Normal Function, Ultrasound Med Biol., 45(5), 1197-1207, (2019).

[19] Andersen, M.V, et al., Feature tracking algorithm for circumferential strain using high frame rate echocardiography, 2016 Computing in Cardiology Conference (CinC), pp. 885-888, (2016).

[20] Andersen, M.V, Moore, C., et al., High-Frame-Rate Deformation Imaging in Two Dimensions Using Continuous Speckle-Feature Tracking, Ultrasound Med Biol., 42(11), 2606-2615, (2016).

[21] Hemmsen, M.C., Petersen, M.M, et al., Ultrasound image quality assessment: a framework for evaluation of clinical image quality, Proc. SPIE 7629, Ultrasonic Imaging, Tomography, and Therapy, 76290C, (2010).

[22] Zhang, S., Wang, Y., et al., CNN-Based Medical Ultrasound Image Quality Assessment, Complexity, vol. 2021, Article ID 9938367, 9 pages, (2021).

[23] Sassaroli, E., Crake, C., et al., Image quality evaluation of ultrasound imaging systems: advanced B-modes, J Appl Clin Med Phys., 20(3), 115–124, (2019).

[24] Rukundo, O., Schmidt, S., Aliasing Artefact Index for Image Interpolation Quality Assessment, Proc. SPIE 10817, Optoelectronic Imaging and Multimedia Technology V, 108171E, (2018).

[25] Rukundo, O., Effects of Improved-Floor Function on the Accuracy of Bilinear Interpolation Algorithm, Computer and Information Science, 8(4), (2015).

[26] Rukundo, O., Schmidt, S., Effects of Rescaling Bilinear Interpolant on Image Interpolation Quality, Proc. SPIE 10817, Optoelectronic Imaging and Multimedia Technology V, 1081715, (2018).

[27] Rukundo, O., Wu, K.N., and Cao, H.Q., Image Interpolation based on the Pixel Value corresponding to the Smallest Absolute Difference, IEEE Int. Workshop on Advanced Computational Intelligence, 432–435, (2011).

[28] Rukundo, O., Normalized Weighting Schemes for Image Interpolation Algorithms, In: ArXiv: 2011.08559, (2020).

[29] Rukundo, O., Pedersen, M., Hovde, Ø., Advanced Image Enhancement Method for Distant Vessels and Structures in Capsule Endoscopy, Computational and Mathematical Methods in Medicine, (2017).

[30] Rukundo, O., Half-Unit Weighted Bilinear Algorithm for Image Contrast Enhancement in Capsule Endoscopy, International Conference on Graphic and Image Processing, 106152Q-1 - 106152Q-9, SPIE, (2018).

[31] Rukundo, O., Effects of Empty Bins on Image Upscaling in Capsule Endoscopy, International Conference on Digital Image Processing, 104202P-1 - 104202P-8, SPIE, (2017).

[32] Ning, Y., Liu, Y., Zhang, Y. et al. Adaptive image rational upscaling with local structure as constraints. Multimed Tools Appl 78, 6889–6911 (2019).

[33] Li, Y., Qi, F., Wan, Y., Improvements On Bicubic Image Interpolation, 4th Advanced Information Technology, Electronic and Automation Control Conference, 1316-1320, (2019).

[34] Khan, S., Lee, D-H., et al., Image Interpolation via Gradient Correlation Based Edge Direction Estimation, Scientific Programming, Volume 2020, Article ID 5763837, 12 pages, (2020).

[35] Sánchez-García, E, Balaguer-Beser, Á., et al., A New Adaptive Image Interpolation Method to Define the Shoreline at Sub-Pixel Level. Remote Sensing., 11(16):1880, (2019).

[36] Interpolated Scan Conversion of B Mode Cardiac Ultrasound Sequences:<https://github.com/orukundo/Interpolated-Scan-Conversion-of-B-Mode-Cardiac-Ultrasound-Image-Sequences >, Accessed: 2021-10-12.